\documentclass[lettersize,journal]{IEEEtran}
\usepackage{amsmath,amsfonts}
\usepackage{algorithmic}
\usepackage{array}
\usepackage{textcomp}
\usepackage{stfloats}
\usepackage{url}
\usepackage{verbatim}
\usepackage{graphicx}
\usepackage{cite}
\usepackage{subcaption}
\usepackage{amssymb}
\usepackage[dvipsnames]{xcolor}
\usepackage{booktabs} 
\usepackage{bm} 
\usepackage[linesnumbered,ruled,vlined]{algorithm2e}
\usepackage{diagbox}
\usepackage{slantsc}
\usepackage{multirow}
\usepackage{cases}
\usepackage{float}
\usepackage{tabto}
\newcommand{\abs}[1]{\left|#1\right|}
\newcommand{\transpose}{^{\mathsf{T}}}
 
\newcommand{\hd}[0]{\mathsf{d_H}} 

\newcommand{\ifconditon}[0]{\mathsf{B}} 
\newcommand{\srrule}[0]{\Phi} 
\newcommand{\bdd}[0]{\mathsf{BDD}} 
\newcommand{\fliplrb}[0]{\textsf{flipLRB}}

\newcommand{\imax}[0]{I_{\max}}

\newcommand{\U}{\mathsmaller{U}}
\newcommand{\W}{\mathsmaller{W}}
\graphicspath{{figure/}}

\begin{document}
	\title{Four-Level Quantized Soft Information Aided Hard-Decision Decoder for Product-like Codes} 
	\author{ 
		\IEEEauthorblockN{ Xiaowei Wu\IEEEauthorrefmark{1}, \IEEEmembership{Member, IEEE}, Min Qiu\IEEEauthorrefmark{2}, \IEEEmembership{Senior Member, IEEE}, Yueying Zhan\IEEEauthorrefmark{1}, Lei Yang\IEEEauthorrefmark{1} }\\
		\IEEEauthorblockA{\IEEEauthorrefmark{1}Technology and Engineering Center for Space Utilization, Chinese Academy of Sciences, Beijing, China \\ 
			Email: \{wuxiaowei,zhanyueying,yang.lei\}@csu.ac.cn}\\
		\IEEEauthorblockA{\IEEEauthorrefmark{2}Global College, Shanghai Jiao Tong University, Shanghai, China. \\Email: min\_qiu@sjtu.edu.cn }
		\thanks{This research was supported in part by the National Natural Science Foundation of China under Grant 62301533, 62671374, 62371439, and U24A20214; in part by the National Key R$\textnormal{\&}$D Program of China under Grant 2025YFF0522600; in part by the Prospective Foundation of CSU-CAS under Grant T4034711QY and T4035511QY. The work of Min Qiu was supported in part by the Shanghai Jiao Tong University ExploreX Funding under Grant SD6040004/153.}
	}
	\IEEEaftertitletext{\vspace{-3\baselineskip}}
	\maketitle
	\begin{abstract}
		We propose a novel low-complexity soft-aided hard-decision decoding scheme for product-like codes, termed iterative bounded-distance decoding (iBDD) with graded scaled reliability and bit-flipping (GSRF). This scheme operates on four-level quantized soft information. It incorporates a graded scaled reliability rule to assign decoding confidence to the bounded-distance decoding result, and employs a bit-flipping rule based on bit-level and component-code-level reliability. A selective re-decoding rule is designed to further facilitate the error performance while keeping the number of BDD attempts in reasonably low levels. The decoding parameter design of iBDD-GSRF is investigated for product codes and staircase codes, respectively. Simulation results on both codes verify that iBDD-GSRF achieves a gain of more than 0.4 dB over the conventional iBDD, comparable to several existing unquantized soft-aided decoders.  
	\end{abstract}
	
	\section{Introduction}
	Product codes (PCs) \cite{productcode} and their variants, such as staircase codes (SCCs) \cite{Smith-staircase,9812617} and zipper codes \cite{Sukmadji-zipper}, have been used for high-speed fiber-optic communications (e.g., 400ZR \cite{ofec}). Recently, soft-aided hard-decision decoding (SA-HDD) schemes \cite{Liga-sabm,Sheikh-ibddsr,Sheikh-ibddcr,Sheikh-beepc,Miao-drsd,Miao-drsd2,Rapp-drsd3,Lei-isabm,Rapp-eed,Zhao-ibddtb,Li-ibddrf} for product-like codes have attracted substantial research interest. In these schemes, the decoder leverages soft channel information while retaining iterative bounded-distance decoding (iBDD) as the core decoding algorithm, aiming to achieve a favorable trade-off between error performance and computational complexity.
	
	Existing SA-HDD schemes have demonstrated significant error performance gain over iBDD, which is the most commonly used hard-decision decoding (HDD) scheme, while maintaining a lower complexity than soft-decision turbo product decoding (TPD) \cite{chase,tpd,Yifei-ofecDec1,Zhongfeng-tpcdec}. Most notably, dynamic reliability score decoding (DRSD) \cite{Miao-drsd} and its extensions \cite{Miao-drsd2,Rapp-drsd3} have shown error performance close to TPD. However, practical implementations of SA-HDDs still face challenges. Advanced schemes like DRSD and message-passing based on error-and-erasure decoding (BEE) \cite{Sheikh-beepc} generally demand high-resolution soft-information quantization. This imposes substantial memory and computation overheads, making them less suitable for resource-limited platforms. In contrast, iBDD with scaled reliability (SR) \cite{Sheikh-ibddsr} and its combined-reliability (iBDD-CR) variant \cite{Sheikh-ibddcr} demonstrate the capability of operating under low quantization levels, offering a more hardware-friendly alternative.
	
	Motivated by this, we propose a low-complexity SA-HDD scheme named ``\emph{iBDD with graded-SR and bit-flipping}'' (iBDD-GSRF). This scheme uses 4-level quantized soft information to attain the memory-efficient advantage and introduces two soft-information-based rules. The graded-SR rule assigns a confidence level to qualify the decoding reliability of each decoding attempt. The bit-flipping rule jointly takes bit-wise channel reliability and component decoding reliability into consideration to select the flipping bits. While having low BDD attempts and memory costs, performance improvements compared to multiple benchmark SA-HDDs are observed. 
	
	\emph{Notations}: 
	Scalars and vectors are written in light and small bold letters, respectively, e.g., $x$ and $\bm{x}$.  $\hd(\bm{x},\bm{y})$ denotes the Hamming distance between binary vectors $\bm{x}$ and $\bm{y}$. $\oplus$ denotes the bit-wise XOR operator. $\ifconditon(\mathsf{cond})$ is a hard-decision operator that outputs either 1 or 0 to indicate whether the given condition ``$\mathsf{cond}$'' is satisfied or not. 
	
	\section{Research Background}
	
	Consider transmitting binary phase shift keying signaling over the additive white Gaussian noise (AWGN) channel. The channel output corresponding to a transmitted bit $x\in \{0,1\}$ is given by $r=(-1)^{1-x} + z$, where $z$ is the real-valued Gaussian noise with zero mean and variance $\sigma^2 = 1/(2 R E_b/N_0)$, $R$ is the code rate, and  $E_b/N_0$ is the signal-to-noise ratio (SNR) per information bit. 
	
	\subsubsection{Bose-Chaudhuri-Hocquenghem (BCH) codes and BDD}
	Let $\mathcal{C}(n, k, t)$ be a BCH code with codeword length $n$, information length $k$, and designed error correction capability $t$. 
	Given a transmitted BCH codeword $\bm{x}\in\mathcal{C}$ and its corresponding received sequence $\bm{y}$, the output of bounded-distance decoding (BDD) can be written as
	\begin{align}\hspace{-2pt} \label{eq:bdd}
		\bdd(\bm{y}) \!=\! \left\lbrace  \begin{aligned}
			& \!\bm{x}&& \!\!\!\text{if } \hd(\bm{x},\bm{y}) \!\leq\! t 
			&& \!\!\text{\footnotesize(valid result)}
			\\
			& \!\tilde{\bm{x}} && \!\!\!\text{if }  \hd(\tilde{\bm{x}},\bm{y}) \!\leq\!  t \!<\! \hd(\bm{x},\bm{y}) 
			&& \!\!\text{\footnotesize(miscorrection)}
			\\
			& \!\bm{y}&&  \!\!\!\text{if } \hd(\bm{y}, \tilde{\bm{x}}) \!>\! t \,\forall\, \tilde{\bm{x}} \!\in\! \mathcal{C}   
			&&\!\!\text{\footnotesize(BDD failure)}
		\end{aligned}
		\right. .
	\end{align}
	In the case of miscorrection, $\tilde{\bm{x}}$ is an erroneous estimation of $\bm{x}$ such that $\tilde{\bm{x}} \neq \bm{x}$ and $\tilde{\bm{x}}\in\mathcal{C}$. 
	
	\subsubsection{Product Codes}
	We focus on a PC with each row and column being a codeword of the identical BCH component $\mathcal{C}$. The PC has a code rate of $R=k^2/n^2$. 
	
	A received PC block is processed with an iterative decoding scheme that alternately decodes the row and column of the block with a component BCH decoder. Typically, intrinsic message passing (IMP) \cite{Rapp-eed} is used so that the output from the row decoding directly becomes the input to the column decoding, and vice versa. The iteration terminates when the decoder either reaches the maximum number of iterations $\imax$ or when all rows and columns have zero syndrome, i.e., each row and column is a valid codeword of $\mathcal{C}$.
	
	\subsubsection{Staircase Codes}	
	An SCC is built up with the set of binary blocks $\mathbf{B}_i$ of size $n/2 \times n/2$, where $i=1,2,\dots$, and $\mathbf{B}_0$ is set for initialization. The $i$-th spatial position of SCC comprises $\mathbf{C}_i=[\mathbf{B}\transpose_{i-1},\mathbf{B}_{i}]$, where each row of $\mathbf{C}_i$ is a codeword of the component BCH code. The code rate of SCC with component codes $\mathcal{C}(n, k, t)$ is given by $R=1-\frac{2(n-k)}{n}$. 
	
	Iterative decoding of staircase codes is performed in a sliding-window manner \cite{Smith-staircase}. Given a window size of $W$ spatial position, the $U$-th window section includes $W+1$ received blocks $\mathbf{Y}_{\U-1}, \dots, \mathbf{Y}_{\U+\W-1}$, corresponding to spatial positions $\mathbf{C}_{\U}, \dots, \mathbf{C}_{\U+\W-1}$. In the $i$-th iteration of this window, the component decoder slides backward from the right end $[\mathbf{Y}\transpose_{\U+\W-2},\mathbf{Y}_{\U+\W-1}]$ to the left end $[\mathbf{Y}\transpose_{\U-1},\mathbf{Y}_{\U}]$ within the window. At the end of this window section, the estimation of $\mathbf{B}_{\U-1}$ is given as the output of the window. 
	
	\section{iBDD With Graded-SR and Bit-Flipping}		
	In the following, we start with iBDD-GSRF for PCs and later extend the approach to SCCs. Following the conventional iBDD decoding scheme reviewed in the last section, our iBDD-GSRF also uses IMP \cite{Rapp-eed} for iterative decoding. 
	
	\begin{figure}[t]
		\centering
		\includegraphics[width=0.95\linewidth]{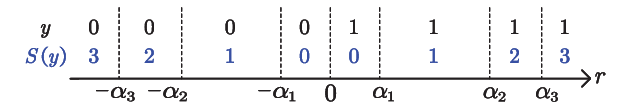}
		\caption{Quantization of the received signal.}
		\label{fig:quantization}
		\vspace{3pt}
		\includegraphics[width=0.95\linewidth]{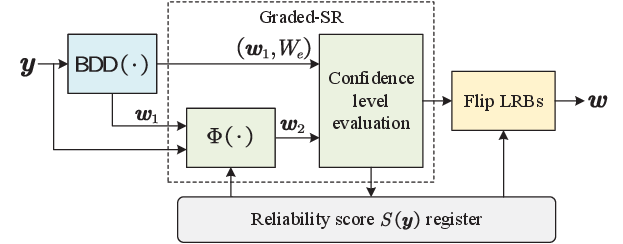}
		\caption{Block diagram of iBDD-GSRF.}
		\label{fig:decoder_block_diagram}
		\vspace{-15pt}
	\end{figure}
	
	\begin{algorithm}[b]
		\caption{BDD with GSRF} \label{alg:sabdd}
		\DontPrintSemicolon
		\tcc{step 1:~\small bounded-distance decoding}
		$(\bm{w}_1,W_e)\leftarrow \mathsf{BDD}(\bm{y})$  \label{ln:bdd} \tcp{see \eqref{eq:bdd} \eqref{eq:bdd_e}}
		\tcc{step 2:~graded-SR rule}
		Apply SR rule: $\bm{w}_2\leftarrow \srrule(\bm{w}_1,\bm{y})$ \label{ln:sr} \tcp{see \eqref{eq:sr_rule}-\eqref{eq:sr_rule_3}} 
		\tcp{Grade the SR decision $\bm{w}_2$} 
		\textbf{if }{$\bm{w}_1 \notin \mathcal{C}$} \tab\hspace{-35pt}\textbf{then} {$V \leftarrow -2$}\; 
		\textbf{else if }{$W_c = W_e$} \tab\hspace{-35pt}\textbf{then} {$V \leftarrow -1$ \label{ln:v_b}}\; 
		\textbf{else if }{$0 < W_c < W_e$} \tab\hspace{-35pt}\textbf{then} {$V \leftarrow 0$}\;
		\textbf{else if }{$W_c = 0$} \tab\hspace{-35pt}\textbf{then} {$V \leftarrow 1$}\; 
		\tcc{step 3:~bit-flipping}
		Update $S(\bm{y})$ with \eqref{eq:s_update}\; \label{ln:v1}
		\textbf{if }{$V\in\{0,1\}$} \textbf{then} {$\bm{w} \leftarrow \bm{w}_2 $ \;}
		\ElseIf{$V\in\{-2,-1\}$}{
			$\bm{w} \leftarrow \fliplrb(\bm{y})$ \label{ln:fliplrb}  \tcp{defined in \eqref{eq:fliplrb}} 
		}
		\tcc{Optional step:~re-decoding }
		\If{re-decoding ON \textnormal{\textbf{and}} $V=-2$}{
			$(\bm{w}_1,W_e)\leftarrow \mathsf{BDD}(\bm{w})$\; 
			Execute lines 2-10, then go to line \ref{ln:return}\;
		}
		\Return{$\bm{y}\leftarrow\bm{w}$} \label{ln:return} \tcp{\small overwrite component input}
	\end{algorithm}
	
	\subsection{Decoding Product Codes}
	Given a channel output $r$, we let ${y=\ifconditon(r>0)}$ be the hard-decision of $r$, and $S(y)\in\{0,1,2,3\}$ be the reliability score of $y$, such that a larger $S(y)$ means a higher probability of $y$ being correct. As shown in Fig. \ref{fig:quantization}, $(\alpha_1,\alpha_2,\alpha_3)$ are the quantization thresholds that satisfy ${0 < \alpha_1 < \alpha_2 < \alpha_3}$, and $S(y)$ is initialized by a non-uniform 4-level quantizer
	\begin{align} 
		S(y)= \textstyle \sum_{i=1}^{3} \ifconditon(\abs{r} > \alpha_i) .
	\end{align} 
	
	The decoding process of  a row/column component $\bm{y}=[y_1,\dots,y_n]$ involves three steps: 
	\begin{enumerate}
		\item Produce BDD result $\bm{w}_1=[w_{1,1},\dots,w_{1,n}]$.
		\item Apply the \emph{graded-SR} (GSR) rule to produce the second-stage decision $\bm{w}_2=[w_{2,1},\dots,w_{2,n}]$.
		\item Apply \emph{bit-flipping} based on the results of steps 1-2 and the bit-wise reliability score $S(\bm{y})=[S(y_1),\dots,S(y_n)]$, producing the final decision $\bm{w}=[w_1,\dots,w_n]$. 
	\end{enumerate}
	Last, $\bm{w}$ directly overwrites the component decoding input $\bm{y}$. The details are given in Alg. \ref{alg:sabdd} and illustrated in Fig. \ref{fig:decoder_block_diagram}.
	
	\subsubsection{BDD decoding (line 1)} 
	In step 1, the BDD decoder inputs a row/column component $\bm{y}$ of the received PC block and outputs the decoding result $\bm{w}_1$. It also outputs the Hamming distance between $\bm{y}$ and $\bm{w}_1$, which indicates the estimated number of bit errors in $\bm{y}$, expressed by
	\begin{align}
		W_e =\hd(\bm{y},\bm{w}_1) = \textstyle \sum_{i=1}^{n} \ifconditon(y_i \neq w_{1,i}) . \label{eq:bdd_e}
	\end{align}
	
	\subsubsection{Graded SR (lines 2-6)} 
	In step 2, we use the SR rule \cite{Sheikh-ibddsr} to identify BDD miscorrection. The SR decision vector $\bm{w}_2$ is produced by applying a bit-wise check to $\bm{w}_1$, where
	\begin{subnumcases}
		{w_{2,i} \!=\! \srrule(w_{1,i},y_i) \!\triangleq\! \label{eq:sr_rule}}
		\! y_i \!\!\!		& if $w_{1,i} \!\neq\! y_i$ and $|r_i| \!\geq\! T_s$ \label{eq:sr_1}\\
		\! w_{1,i}\!\!\!	& if $w_{1,i} \!\neq\! y_i$ and $|r_i| \!<\! T_s$\\
		\! w_{1,i}\!\!\!	& if $w_{1,i} \!=\! y_i$ 	 	
	\end{subnumcases}
	for $i=1, \dots, n$, where $T_s$ is the SR decision threshold and will be determined using \eqref{eq:sr_rule_2}. 
	
	The core idea of the SR rule \cite{Sheikh-ibddsr} is to leverage channel reliability, indicated by the magnitude of the received signal $|r_i|$, to reject possible miscorrection. When $|r_i| \geq T_s$, the SR operator identifies the hard-decision $y_i=\ifconditon(r_i)$ as an anchor bit that is assumed to be correct. In this case, if the BDD decision $w_{1,i}$ conflicts with $y_i$, the SR operator identifies $w_{1,i}$ as a miscorrected bit and rejects it by forcing $w_{2,i}=y_i$. To match the decrease in BDD miscorrection probability across iterations, the original SR rule monotonically increases $T_s$, requiring $2I_{\max}$ distinct threshold levels.
	
	To fit the 4-level quantization, our scheme simplifies the updating of $T_s$ into a two-stage step function, given by	
	\begin{align} 
		\label{eq:sr_rule_2}
		T_s=\left\{\begin{aligned}
			&2 && \textrm{for } I=1,\dots,\imax/2\\
			&3 && \textrm{for } I=\imax/2+1,\dots,\imax
		\end{aligned} \right. . 
	\end{align}
	The anchor condition $|r_i| \geq T_s$ simplifies to $S(y_i) \geq T_s$. Following \eqref{eq:sr_1}, the number of conflicts between the BDD decision and anchor bits is measured by the Hamming distance between $\bm{w}_1$ and $\bm{w}_2$:
	\begin{align} \label{eq:sr_rule_3}
		W_c \!=\! \hd(\bm{w}_1,\bm{w}_2) \!=\! \textstyle \sum_{i=1}^{n} \ifconditon(w_{1,i} \!\neq\! y_i \textrm{ and }S(y_i) \!\geq\! T_s) .
	\end{align}%
	
	To compensate for the decision ambiguity caused by the low-resolution SR operation, we introduce a confidence level ${V\in\{-2,-1,0,1\}}$ for $\bm{w}_2$ to quantify its global reliability. Upon \emph{BDD failure}, the lowest level $V=-2$ is assigned, meaning that $\bm{w}_2$ does not form a valid BCH codeword. Otherwise, $V$ is assigned by tracking the conflicts between BDD decision $\bm{w}_1$ and anchor bits:
	\begin{itemize}
		\item \emph{Identified miscorrection} (${V\!=\!-1}$): All corrections of BDD conflict with anchor bits (${W_c = W_e}$), indicating a high likelihood of BDD miscorrection. To prevent error propagation, the SR decision $\bm{w}_2$ completely reverts to $\bm{w}_2 = \bm{y}$.
		\item \emph{Tentative decoding} (${V\!=\!0}$): Only a portion of BDD corrections conflict with anchor bits (${0 \!<\! W_c \!<\! W_e}$), indicating partial uncertainty. The SR decision $\bm{w}_2$ accepts non-conflicting corrections while rejecting conflicting ones.
		\item \emph{Confident decoding} (${V\!=\!1}$): No BDD corrections conflict with anchor bits (${W_c\!=\!0}$), indicating high decoding reliability. Thus, the SR decision  $\bm{w}_2$ accepts all BDD corrections, producing $\bm{w}_2=\bm{w}_1$.			
	\end{itemize}	
	
	\emph{\textbf{Example:}} Consider component length $n=8$ and decision threshold $T_s=2$. Suppose $\bm{y}$ and $S(\bm{y})$ are initialized as:	
	\begin{table}[H]
		\vspace{-7pt}
		\centering
		\renewcommand{\arraystretch}{1.25}
		\setlength{\tabcolsep}{3.5pt}
		\begin{tabular}{|c|c|c|}
			\hline
			\textbf{Hard-decision} $\bm{y}$ & \textbf{Reliability scores} $S(\bm{y})$ & \textbf{Threshold} $T_s$ \\ \hline
			$[0, 0, 1, 1, 0, 0, 1, 0]$ & $[2, 3, 0, 1, 2, 1, 1, 0]$ & $2$ \\ \hline
		\end{tabular}
		\vspace{-10pt}
	\end{table}
	\noindent Given these conditions, the GSR decisions for different BDD outputs are given below:	
	\begin{table}[H]
		\vspace{-7pt}
		\centering
		\renewcommand{\arraystretch}{1.25}
		\setlength{\tabcolsep}{2pt}
		\resizebox{\linewidth}{!}{%
			\begin{tabular}{|c|c|c|c|c|c|c|}
				\hline
				& \textbf{BDD Output} $\bm{w}_1$ & $W_e$ & \textbf{SR Decision} $\bm{w}_2$ & $W_c$ & \textbf{Condition} & $V$ \\ \hline
				(i)& $[\underline{\mathbf{1}}, \underline{\mathbf{1}}, 1, 1, \mathbf{0}, 0, 1, 0]$ & 2 & $[\mathbf{0}, \mathbf{0}, 1, 1, \mathbf{0}, 0, 1, 0]$ & 2 & $W_c \!=\! W_e$ & $-1$ \\ \hline
				(ii)& $[\underline{\mathbf{1}}, \mathbf{0}, \underline{0}, 1, \mathbf{0}, 0, 1, 0]$ & 2 & $[\mathbf{0}, \mathbf{0}, \underline{0}, 1, \mathbf{0}, 0, 1, 0]$ & 1 & $0 \!<\! W_c \!<\! W_e$ & $0$ \\ \hline
				(iii)& $[\mathbf{0}, \mathbf{0}, \underline{0}, \underline{0}, \mathbf{0}, 0, 1, 0]$ & 2 & $[\mathbf{0}, \mathbf{0}, \underline{0}, \underline{0}, \mathbf{0}, 0, 1, 0]$ & 0 & $W_c \!=\! 0$ & $1$ \\ \hline
				\multicolumn{7}{|p{8.9cm}|}{\footnotesize *Underline denotes the bit-position where a BDD correction ($w_{1,i} \!\neq\! y_i$) appears, and boldface denotes the bit-position of anchor bits ($S(y_i) \!\geq\! T_s$).}\\ \hline
			\end{tabular}%
		}
		\vspace{-10pt}
	\end{table}
	In case-i, BDD flips $y_1$ and $y_2$. Since both bits meet the anchor condition, all corrections are rejected, yielding $V=-1$. In case-ii, BDD flips $y_1$ and $y_3$. Since only $y_1$ is an anchor bit, $w_{1,1}$ is rejected while $w_{1,3}$ is accepted, returning $V=0$. In case-iii, BDD flips $y_3$ and $y_4$. Neither correction conflicts with anchor bits, so all BDD corrections are accepted. 
	\hfill $\blacksquare$
	
	\subsubsection{Bit-flipping (lines 7-10)} 
	In Step 3, depending on the confidence level $V$ of the SR decision $\bm{w}_2$, the decoder accepts the decision or applies bit-flipping. For a tentative $(V=0)$ or confident $(V=1)$ result, $\bm{w}_2$ is directly outputted. Upon BDD failure ${(V=-2)}$ or identified miscorrection ${(V=-1)}$, the least reliable bits (LRBs) in $\bm{y}$ are randomly flipped to accelerate convergence \cite{Liga-sabm}. To be specific, a bit $y_{i}$ in $\bm{y}$ is identified as an LRB if it satisfies the following conditions: 
	\begin{itemize}
		\item \emph{Low channel reliability}: its channel magnitude satisfies $|r_{i}|\leq\alpha_1$, such that its reliability score is initialized as $S(y_i)=0$;
		\item \emph{Low decoding reliability}: up to current decoding attempt, $y_{i}$ has never achieved a confident decoding result (${V\!=\! 1}$).
	\end{itemize}
	\noindent To track bits with \emph{low decoding reliability} without additional memory overhead, the reliability score is updated by
	\begin{align} \label{eq:s_update}
		S(y_i) =\left\{\begin{aligned}
			&\max\!\big(S(y_i),1\big) && \textrm{if }V=1\\
			&S(y_i) && \textrm{otherwise}
		\end{aligned} \right. ,
	\end{align}
	for $i=1,\dots,n$. This prevents bits in a highly reliable row/column decoding result from being wrongly flipped in the succeeding decoding attempts, thereby suppressing decoding oscillation and accelerating convergence. Hence, the bit-flipping operation can be expressed as 
	\begin{align} \label{eq:fliplrb}
		w_i =\left\{\begin{aligned}
			&\mathrm{rand}\{0,1\}&& \textrm{if }S(y_i)=0\\
			&y_i && \textrm{otherwise}
		\end{aligned} \right. ,
	\end{align}
	where $\mathrm{rand}\{0,1\}$ returns a random bit value.
	
	To avoid the computation overhead of random bit generation, we introduce a deterministic fixed-pattern bit-flipping rule. Specifically, in the $I$-th decoding iteration, LRBs with odd indices $\big\{y_i|S(y_i)=0,i\in\{1,3,5,\dots,n-1\}\big\}$ are flipped when $I$ is odd, and even-indexed LRBs are flipped when $I$ is even. For re-decoding, the LRBs are flipped using the inverse pattern. Other combinations are also possible.
	
	\begin{figure*}[t]
		\centering
		\begin{minipage}{0.63\textwidth}
			\centering
			\includegraphics[height=4cm]{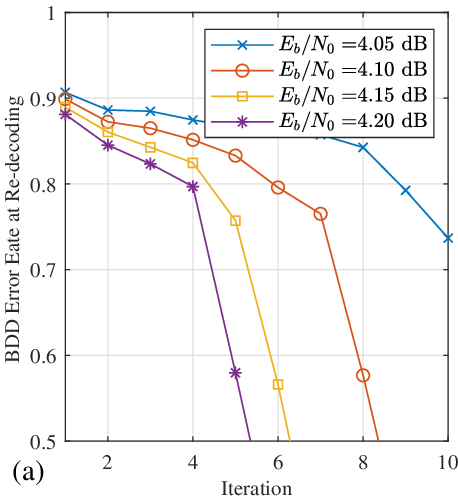}	
			\includegraphics[height=4cm]{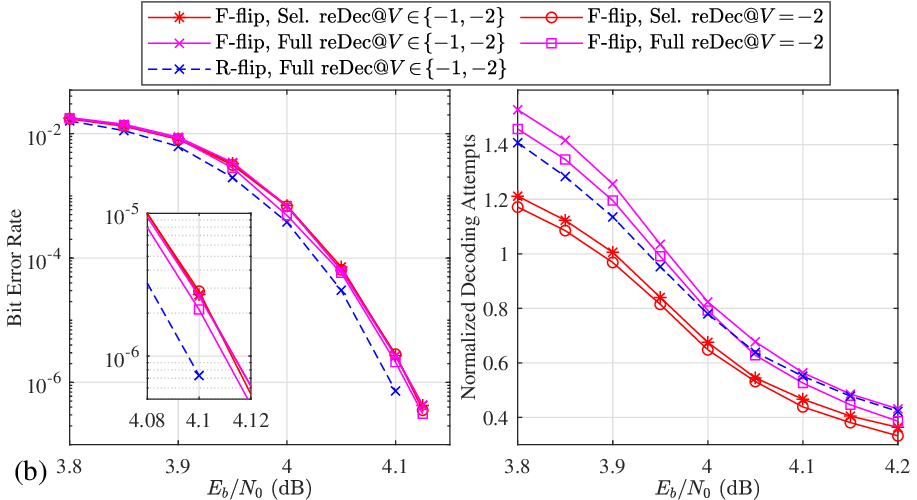}
			\caption{Performance of iBDD-GSRF for PC with $\mathcal{C}_2$ and $\imax=10$. (a) Per-iteration BDD error rate at the re-decoding stage. (b) BER and decoding attempts (normalized by ${255 \!\times\!20}$) for different configurations.} \label{fig:setting_compare}
		\end{minipage}
		\hspace{5pt}
		\begin{minipage}{0.34\textwidth}
			\centering
			\vspace{-20pt}		
			\captionof{table}{BCH component codes and the quantization thresholds for iBDD-GSRF.}\vspace{-3pt}
			\label{tb:codes}
			\setlength{\tabcolsep}{4pt} 
			\resizebox{\linewidth}{!}{%
				\begin{tabular}{lll}
					\toprule
					& Component & Thresholds \\
					\midrule
					PC & $\mathcal{C}_1(255,239,2)$ & $(0.11, 0.4, 0.5)$ \\
					& $\mathcal{C}_{1,e}(255,238,2)$ & $(0.11, 0.4, 0.5)$ \\
					& $\mathcal{C}_2(255,231,3)$ & $(0.1, 0.5, 0.6)$ \\
					& $\mathcal{C}_3(511,484,3)$ & $(0.08, 0.4, 0.5)$ \\
					\midrule
					SCC & $\mathcal{C}_{2,s}(254,230,3)$ & $(0.12, 0.35, 0.45)$\\
					\bottomrule
				\end{tabular}
			}
			\begin{flushleft} 
				\vspace{-2pt}{\footnotesize *$\mathcal{C}_{1,e}(255,238,2)$ is the even-weight subcode of $\mathcal{C}_1(255,239,2)$, and $\mathcal{C}_{2,s}(254,230,3)$ is the 1-bit shorten code of $\mathcal{C}_2(255,231,3)$.} \vspace{-15pt} \end{flushleft}
		\end{minipage}
		\vspace{-15pt}
	\end{figure*}
	
	\subsubsection{Re-Decoding (lines 11-13)} 
	Optionally, the decoder can enable a re-decoding mechanism that repeats steps 1-3 by feeding back the output of step 3 to step 1. When this option is enabled, an actual re-decoding attempt is performed exclusively upon a BDD failure ${(V=-2)}$, rather than an identified miscorrection ${(V=-1)}$. In addition, the decoder revokes re-decoding for the first three iterations. This selective schedule is motivated by the varying re-decoding success rates under different operating conditions.
	
	Specifically, it is observed via simulation that the probability of producing the correct codeword via immediate re-decoding upon ${V=-1}$ is low. Thus, revoking re-decoding for ${V=-1}$ effectively reduces the number of BDD attempts without significantly sacrificing the error performance. In addition, the success probability of re-decoding is inherently limited during early iterations due to high residual error $\hd(\bm{y},\bm{x})$. As depicted in Fig. \ref{fig:setting_compare}a, when all iterations allow re-decoding, the per-iteration BDD error rate at re-decoding is substantially higher for early iterations. Accordingly, the decoder can revoke re-decoding for the first few iterations.  
	
	To evaluate the proposed re-decoding schedule, Fig. \ref{fig:setting_compare}b compares the decoding performance under various flipping patterns and re-decoding schedules:
	\begin{itemize}
		\item ``R-flip''/``F-flip'': random/fixed flipping pattern.
		\item ``Full reDec'': allow re-decoding in all iterations. 
		\item ``Sel. reDec'': revoke re-decoding for the first three iterations, and allow re-decoding for the rest of the iterations. 
		\item ``@${V \!\in\! \{-1,-2\}}$'': re-decoding upon ${V \!=\! -1}$ or ${V \!=\! -2}$.
		\item ``@${V=-2}$'': re-decoding only upon $V=-2$.
	\end{itemize}
	As shown, the proposed re-decoding schedule ``Sel. reDec@${V=-2}$'' shows similar bit error rate (BER) to ``Full reDec@${V \!\in\! \{-1,-2\}}$'', while drastically reducing the decoding attempts. Furthermore, the fixed flipping pattern introduces a negligible loss of less than 0.03 dB compared to random flipping.
	
	\emph{\textbf{Remark:}} The core distinction of iBDD-GSRF over existing SR families \cite{Sheikh-ibddsr,Sheikh-ibddcr,Zhao-ibddtb,Li-ibddrf} lies in the dimension of decision. While conventional SR families rely on bit-wise decisions, we introduce new component-wise decision operations to compensate for the loss of granularity in the bit-wise decisions. By exploiting the inherent constraints of the component BCH codes, the proposed scheme fully utilizes the 4-level quantized soft information while having low complexity. $\hfill\blacksquare$%
	\vspace{-5pt}
	\subsection{Extension to Decoding Staircase Codes}		
	We now extend iBDD-GSRF to SCCs. The high-level iterative decoding architecture still sticks to the conventional SCC sliding-window schedule. The decoding parameters, including the component decoder (hard or soft-aided), the threshold $T_s$, and the re-decoding activation, are configured based on the current iteration and the relative spatial position within the window, as detailed below: 
	\begin{table}[H]
		\centering
		\vspace{-5pt}
		\label{tb:dec_config}
		\setlength{\tabcolsep}{4pt} 
		\renewcommand{\arraystretch}{1.3} 
		\begin{tabular}{|c|c|c|}
			\hline
			\diagbox{Iteration}{Position} & $1,\dots, \lfloor W/2 \rfloor$ & $\lfloor W/2 \rfloor+1,\dots,W$  \\
			\hline
			$1,\dots,\imax\!-\!2$ & GSRF$(T_s=3)$       &  \multirow{2}{*}{GSRF$(T_s=2)$}\\ 
			\cline{1-2}
			$\imax\!-\!1,\imax$ & BDD only    &    \\
			\hline
			\hline
			\diagbox{Iteration}{Position} & $1,\dots,W\!-\!2$ & $W\!-\!1,W$  \\
			\hline
			$1,\dots,\imax$ & re-decoding ON & re-decoding OFF  \\ 
			\hline
		\end{tabular}%
		\vspace{-5pt}
	\end{table}
	
	The parameter design of GSRF for SCCs shares a similar strategy to that for PCs, which is to match the parameters to the number of decoding attempts. As the decoding window slides forward, blocks at leftward positions within the window have completed more decoding attempts than blocks at rightward positions, thereby exhibiting fewer residual errors. Therefore, a strict threshold is applied to the right-half positions and subsequently loosened for the left-half positions. Likewise, re-decoding is initially revoked for the rightmost positions, as high residual errors in newly entered blocks lead to frequent re-decoding failures. Standard BDD is applied exclusively for the leftmost positions in the final iterations, as executing BDD on components with excessive residual errors barely yields valid corrections.
	
	\begin{figure*}[t]
		\centering
		\begin{minipage}{0.66\textwidth}
			\centering
			\subcaptionbox{$\mathcal{C}_1,\mathcal{C}_2,\mathcal{C}_3$ and $\imax=10$} {\includegraphics[height=4.5cm]{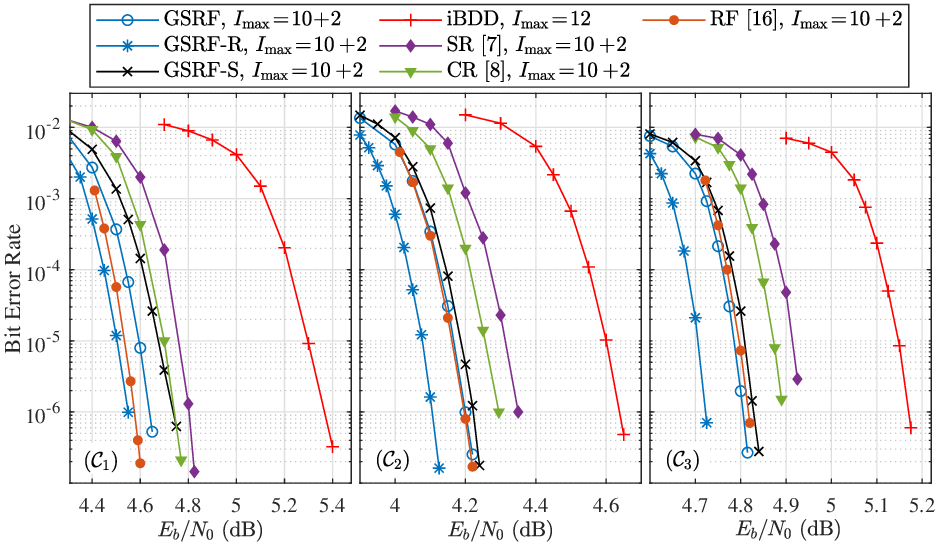}}%
			\hspace{3pt}
			\subcaptionbox{$\mathcal{C}_{1,e}$ and $\imax=10,20$} {\includegraphics[height=4.5cm]{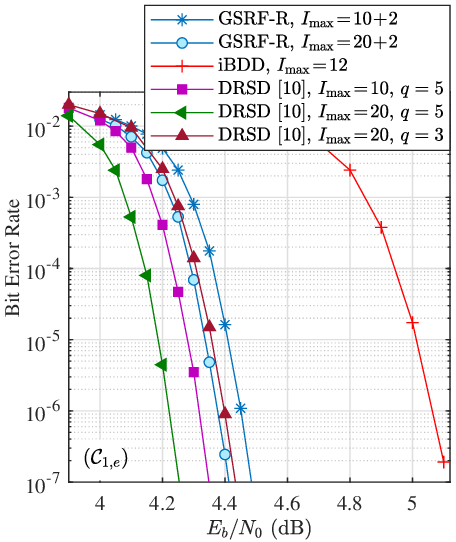}}%
			\vspace{-3pt}
			\caption{BER performance of SA-HDDs for PCs with $\mathcal{C}_1,\mathcal{C}_{1,e},\mathcal{C}_2,\mathcal{C}_3$.} \label{fig:BER_C123}
		\end{minipage}
		\begin{minipage}{0.32\textwidth}
			\centering%
			\includegraphics[height=4.5cm]{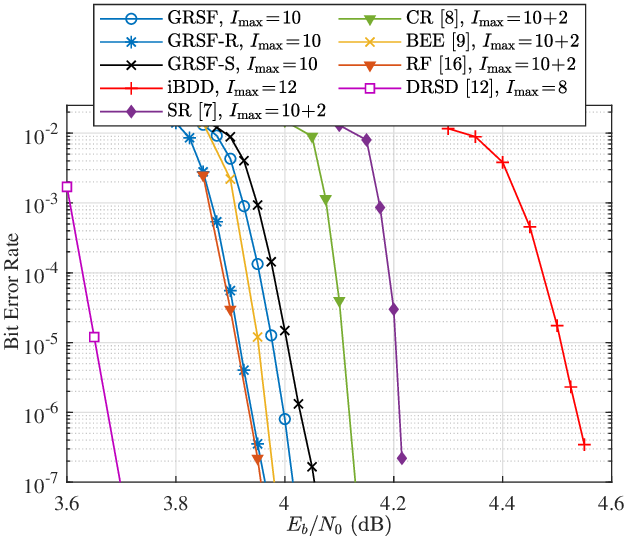}
			\caption{BER performance of SA-HDDs for SCC with 1-bit shortened BCH $\mathcal{C}_{2,s}$.} \label{fig:BER_pc_scc}
		\end{minipage}
		\vspace{-15pt}
	\end{figure*}
	
	\section{Numerical Results and Discussions}
	We evaluate the performance of iBDD-GSRF using Monte-Carlo simulations. The BCH codes used for constructing PCs and SCCs are summarized in Table \ref{tb:codes}. The quantization thresholds ${(\alpha_1, \alpha_2, \alpha_3)}$ for iBDD-GSRF are optimized for each code ensemble and component code, and are fixed regardless of signal-to-noise ratio (SNR). The constraint ${\alpha_3=\alpha_2+0.1}$ is set to simplify the threshold design. Then, $(\alpha_1,\alpha_2)$ is optimized using a numerical search. The quantization thresholds used for simulations are listed in Table \ref{tb:codes}. Notably, slight variations in the quantization thresholds ($\pm 0.05$) do not severely degrade the performance of the considered codes.
	
	For the simulations of iBDD-GSRF, the fixed flipping pattern is adopted, and three variants are evaluated:
	\begin{itemize}
		\item \emph{GSRF}: no re-decoding,
		\item \emph{GSRF-R}: with the proposed selective re-decoding,
		\item \emph{GSRF-S}: with ${T_s\!=\!2}$ fixed, and no re-decoding.
	\end{itemize}  
	For decoding PCs, two iBDD iterations are performed after $\imax$ iBDD-GSRF iterations. For decoding SCCs, no additional iBDD iteration is executed.
	
	We consider iBDD-SR \cite{Sheikh-ibddsr}, iBDD-CR \cite{Sheikh-ibddcr}, iBDD-RF \cite{Li-ibddrf}, BEE-SCC\cite{Sheikh-beepc}, and DRSD\cite{Miao-drsd,Rapp-drsd3} as benchmarks. All benchmark results are reported in their respective original papers. Except for DRSD, which employs $q$-bit reliability score ($q=5$ by default) and $1$-bit erasure marker, all other benchmarks use unquantized soft information. DRSD employs only $\imax$ soft-aided iterations, whereas the remaining benchmarks perform $\imax$ soft-aided iterations followed by two iBDD iterations.
	
	Figs. \ref{fig:BER_C123}-\ref{fig:BER_pc_scc} show the BER performance for PCs and SCCs, respectively. Compared to iBDD, iBDD-GSRF achieves a significant gain of 0.3-0.75 dB. Notably, iBDD-GSRF outperforms iBDD-SR/CR and has comparable performance to iBDD-RF and the DRSD with $q=3$. While adopting a fixed threshold $T_s=2$ incurs a slight performance loss compared to the two-stage $T_s$ schedule, iBDD-GSRF can still outperform iBDD-SR/CR.
	
	Compared to iBDD, the primary complexity overhead of SA-HDDs is the increased BDD attempts. To evaluate this, Fig. \ref{fig:N255K239_decNum} plots the average number of BDD attempts, normalized by ${255 \!\times\!20}$, for iBDD-GSRF and benchmark SA-HDDs \cite{Sheikh-ibddsr,Li-ibddrf,Miao-drsd}. At high SNRs, iBDD-GSRF has fewer decoding attempts than iBDD-SR/RF and DRSD, while achieving error performance comparable to iBDD-RF. 
	
	\begin{figure}[t]
		\centering
		\vspace{-3pt}
		\includegraphics[width=6.5cm]{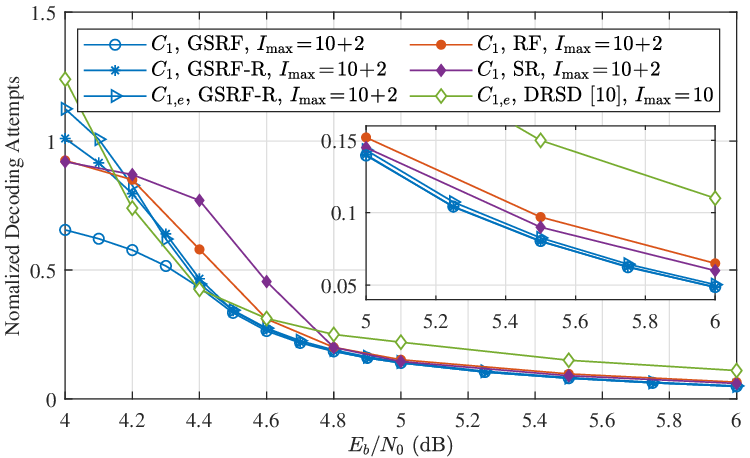}\vspace{-3pt}
		\caption{Decoding attempts for PCs at $\imax=10$.} 
		\label{fig:N255K239_decNum}
		\vspace{-15pt}
	\end{figure}
	
	Regarding the additional memory costs compared to iBDD, iBDD-GSRF requires $2n^2$ bits of memory to store soft information. Two bit-vectors of length $\imax$ are used for indicating $T_s$ and the selective re-decoding, respectively. Table \ref{tb:memorycost} summarizes the memory costs of iBDD-GSRF and benchmark schemes, where the benchmark results are taken directly from their respective original papers. Although the quantization of iBDD-RF has not yet been investigated, it can be inferred that at least 2-bit quantization is required, as it introduces one more soft-aided step based on iBDD-SR. As one can see, iBDD-GSRF has lower memory costs than benchmark schemes except iBDD-SR. 
	
	\begin{table}[h]
		\centering\vspace{-5pt}
		\caption{Additional memory costs of SA-HDDs for PC.} \vspace{-5pt}
		\label{tb:memorycost}
		\setlength{\tabcolsep}{2pt} 
		\begin{tabular}{l c c c}
			\toprule
			&  & Quantization &  Quantization loss\\
			Scheme & Memory costs & bit-width & verified for PC\\
			\midrule
			iBDD-GSRF & $qn^2 \!+\! 2\imax $       & $q=2 $	 	& - \\
			iBDD-SR  & $q(n^2 \!+\! 2\imax )$     & $q\geq 1$  	& 0.06 dB@$q\!=\!1$, $\mathcal{C}_2$ \cite{Fougstedt-ibddsr_ofc}\\
			iBDD-CR  & $q(n^2 \!+\! 6\imax )$     & $q\geq 3$ 	& 0.07 dB@$q\!=\!3$, $\mathcal{C}_2$ \cite{Sheikh-ibddcr}\\
			iBDD-RF  & $q(n^2 \!+\! 4\imax )$     & -	& - \\
			DRSD  & $(q+1)n^2$  & $3 \!\leq\! q \!\leq\! 6$	& 0.18 dB@$q\!=\!3$, $\mathcal{C}_{1,e}$ \cite{Miao-drsd}\\
			\bottomrule
		\end{tabular}
		\vspace{-15pt}
	\end{table}
	
	\section{Conclusions}
	The proposed iBDD-GSRF scheme, which operates on 4-level quantized soft information,  leverages two-dimensional reliability metrics (bit-wise and component-wise) to investigate the low-granularity soft information. While having low complexity overhead compared to many existing SA-HDDs, it provides a substantial error performance gain over iBDD and several SA-HDDs with unquantized soft information. This makes iBDD-GSRF a promising candidate for optical communication systems used in resource-constrained platforms. \vspace{-3pt}
	
	\bibliographystyle{ieeetr}
	\bibliography{references}
	
\end{document}